\documentclass[sigconf]{acmart}
\AtBeginDocument{%
  }

\copyrightyear{2026}
\acmYear{2026}
\setcopyright{cc}
\setcctype{by}
\acmConference[FPGA '26]{Proceedings of the 2026 ACM/SIGDA International Symposium on Field Programmable Gate Arrays}{February 22--24, 2026}{Seaside, CA, USA}
\acmBooktitle{Proceedings of the 2026 ACM/SIGDA International Symposium on Field Programmable Gate Arrays (FPGA '26), February 22--24, 2026, Seaside, CA, USA}
\acmPrice{}
\acmDOI{10.1145/3748173.3779202}
\acmISBN{979-8-4007-2079-6/2026/02}

\usepackage{listings}
\usepackage{multirow}
\usepackage{tabularx}
\usepackage{booktabs, tabularx, makecell}
\usepackage{array}
\usepackage{graphicx}
\usepackage{subcaption}
\usepackage{tikz}
\usepackage{xspace}
\begin{document}

\title{KANEL\'{E}: Kolmogorov–Arnold Networks for Efficient LUT-based Evaluation}

\author{Duc Hoang}
\authornote{Both authors contributed equally to this research.}
\email{dhoang@mit.edu}
\orcid{0000-0002-8250-870X}
\affiliation{
  \institution{Massachusetts Institute of Technology}
  \city{Cambridge}
  \state{MA}
\country{USA}
}

\author{Aarush Gupta}
\authornotemark[1]
\orcid{0000-0002-0177-4855}
\email{aarushg@mit.edu}
\affiliation{%
  \institution{Massachusetts Institute of Technology}
  \city{Cambridge}
  \state{MA}
\country{USA}
}

\author{Philip Harris}
\email{pcharris@mit.edu}
\orcid{0000-0001-8189-3741}
\affiliation{%
  \institution{Massachusetts Institute of Technology}
  \city{Cambridge}
  \state{MA}
\country{USA}
}

\renewcommand{\shortauthors}{Duc Hoang, Aarush Gupta, and Philip Harris}
\newcommand{\kanele}{KANEL\'{E}\xspace}

\begin{abstract}
Low-latency, resource-efficient neural network inference on FPGAs is essential for applications demanding real-time capability and low power.
Lookup table (LUT)-based neural networks are a common solution, combining strong representational power with efficient FPGA implementation.
In this work, we introduce \kanele, a framework that exploits the unique properties of Kolmogorov–Arnold Networks (KANs) for FPGA deployment.
Unlike traditional multilayer perceptrons (MLPs), KANs employ learnable one-dimensional splines with fixed domains as edge activations, a structure naturally suited to discretization and efficient LUT mapping.
We present the first systematic design flow for implementing KANs on FPGAs, co-optimizing training with quantization and pruning to enable compact, high-throughput, and low-latency KAN architectures.
Our results demonstrate up to a 2700x speedup and orders of magnitude resource savings compared to prior KAN-on-FPGA approaches.
Moreover, \kanele matches or surpasses other LUT-based architectures on widely used benchmarks, particularly for tasks involving symbolic or physical formulas, while balancing resource usage across FPGA hardware.
Finally, we showcase the versatility of the framework by extending it to real-time, power-efficient control systems.
\end{abstract}



\begin{teaserfigure}
  \includegraphics[width=\textwidth]{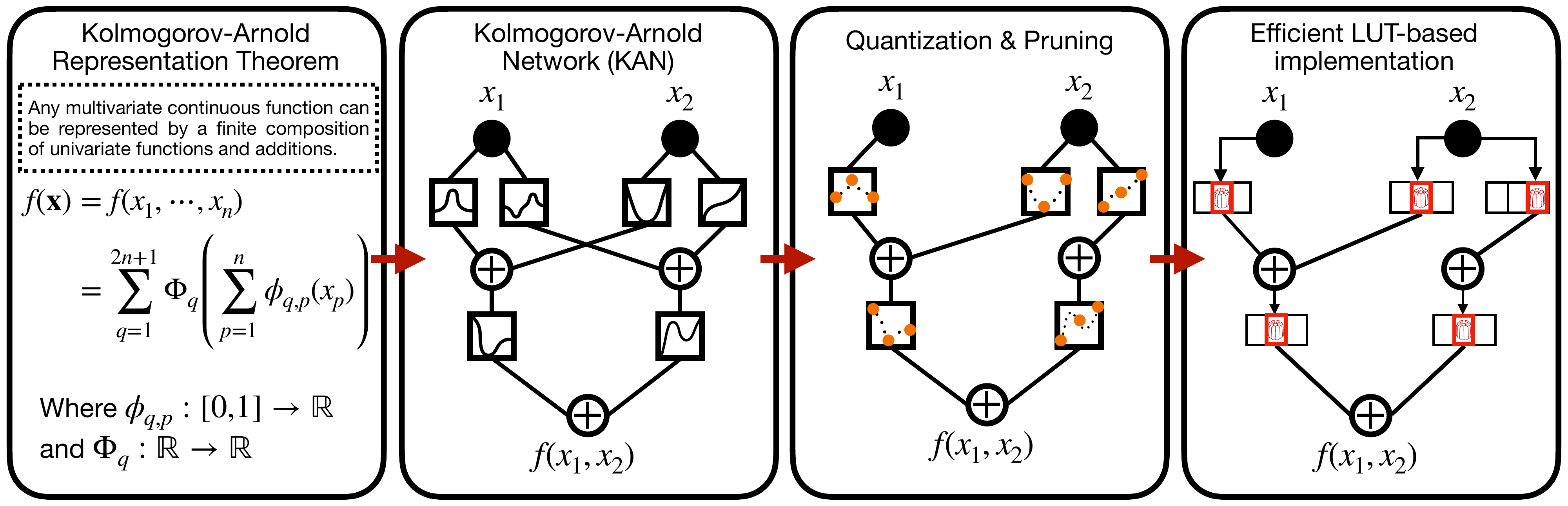}
  \caption{From the Kolmogorov-Arnold Representation Theorem to efficient KAN FPGA inference.}
  \label{fig:teaser}
\end{teaserfigure}

\maketitle

\section{Introduction}

Lookup table (LUT) based neural networks have become a central paradigm for efficient FPGA inference, with designs such as NeuralLUT-Assemble~\cite{NeuralLUT_Assemble}, TreeLUT~\cite{TreeLUT}, DWN~\cite{DWN}, and others~\cite{PolyLUT, AmigoLUT, HGQ, da4ml} demonstrating dramatic gains in area, latency, and power efficiency. 
These approaches highlight the advantages of rethinking neural networks around LUT primitives, though they remain largely confined to supervised learning and task-specific architectures.

In this work, we demonstrate that Kolmogorov–Arnold Networks (KANs) offer a principled foundation for LUT-based design.
Inspired by the Kolmogorov--Arnold representation theorem, KANs replace the fixed activations of Multilayer Perceptrons (MLPs) with learnable edge functions and the matrix multiplication in MLPs with summation at nodes (Fig.~\ref{fig:teaser}). 
This activation-centric formulation aligns naturally with LUTs: each learnable spline defined on a fixed domain can be quantized, pruned, and directly mapped to LUTs. 
In the literature, although KANs have been shown to outperform MLPs in settings such as PDE solving and scientific computing~\cite{KAN, KAN2}, their practical deployment has been hindered by slow inference and costly hardware realizations \cite{Tran_KAN_bad, KAN_BAD_2}. 
The only prior FPGA implementation concluded KANs were \emph{impractical}, due to expensive spline evaluations and high resource usage \cite{Tran_KAN_bad}.

This paper shows that by re-formulating KAN inference entirely in terms of LUTs, KANs are not only feasible but highly efficient in FPGA settings.
Thus, our contributions are fourfold:
\begin{enumerate}

    \item \textbf{FPGA-tailored KAN Architecture:} We present \textsc{\kanele}, named after the French pastry known for its compact form and rich structure \cite{canele}. At its core, the framework co-optimizes quantization, pruning, and mapping of KAN functions onto learned LUTs and additions, thereby minimizing memory and logic overhead.  
    From a KAN research perspective, \textsc{\kanele} is the first FPGA-tailored formulation, eliminating BRAM/DSP usage, reducing latency by up to $2700\times$, and cutting resource usage by over $4000\times$ compared to prior designs~\cite{Tran_KAN_bad}.

    \item \textbf{High-Performance Realizations:}
    Unlike conventional LUT-based neural networks, where sequential LUT indexing makes pruning fundamentally incompatible with the model structure, \textsc{\kanele} leverages the additive independence of KANs to make pruning both natural and hardware efficient.
    Building on this architecture, \textsc{\kanele} delivers FPGA implementations that match or surpass other LUT-based neural designs, particularly for tasks well-suited to symbolic mapping. It sustains clock frequencies above 800\,MHz across most benchmarks while achieving a state-of-the-art Area$\times$Delay product and maintaining a balanced resource footprint.

    \item \textbf{Open-source Framework:} We provide an automated software–hardware co-design flow that compiles KANs into optimized FPGA implementations within seconds, supporting reproducible studies across domains such as biology, physics, vision, signal processing, and tabular ML. Code is available at: \url{https://github.com/Duchstf/KANELE}

    \item \textbf{Control Systems:} we extend \textsc{\kanele} beyond supervised learning to continuous control, showing on the \texttt{HalfCheetah} benchmark from OpenAI Gym \cite{Mujoco} that a quantized KAN policy with $\sim5\times$ fewer parameters than an MLP baseline policy achieves higher rewards, underscoring its suitability for resource-constrained, real-time control systems.

\end{enumerate}

\section{Background \& Related Works}
This section reviews Kolmogorov--Arnold Networks (KANs) and prior work on LUT-based neural network inference.

\subsection{Kolmogorov--Arnold Networks}

Kolmogorov--Arnold Networks (KANs) replace the fixed activation functions and matrix multiplications of MLPs with learnable spline-based functions on network edges~\cite{KAN}. This activation-centric formulation improves expressiveness and interpretability, often achieving comparable accuracy with fewer parameters and operations. Since their introduction, KANs have inspired extensive follow-up work, including theoretical analyses~\cite{kanormlp2024, expressive2025}, architectural extensions (e.g., convolutional~\cite{convkan2025,convprinciples2025}, temporal~\cite{TKAN}, and Fourier-based variants~\cite{kafn2025,FourierKAN2024}), and applications across scientific modeling and data-driven tasks~\cite{Wang_2025,Cruz_2025,li2024ukan,kan4tsf2024}.

Despite this rapid progress, recent surveys identify computational efficiency and hardware implementation as key open challenges~\cite{survey2024, survey2025, KAN_BAD_2}.
To date, efficient hardware realization remains largely unexplored, with one early attempt concluding that a direct FPGA implementation incurs prohibitive resource and latency costs compared to MLPs~\cite{Tran_KAN_bad}.
Our work directly challenges this conclusion by demonstrating that the activation-centric design of KANs is, in fact, exceptionally well-suited for hardware acceleration through a LUT-based paradigm.

\subsection{LUT-based Neural Networks}

LUT-based neural networks aim to replace arithmetic-heavy MAC operations with precomputed function evaluations stored in LUTs, exploiting the abundant and low-latency logic resources of FPGAs.
Pioneering frameworks like LUTNet~\cite{LUTNet} and LogicNets~\cite{LogicNets} first demonstrated the replacement of arithmetic with direct LUT mappings.
Subsequent works generalized this concept to approximate more complex functions~\cite{PolyLUT, NeuraLUT} and improved scalability using additive or modular ensembles~\cite{PolyLut_Add, AmigoLUT, NeuralLUT_Assemble}.
This design philosophy has also been used to efficiently implement other machine learning models, such as gradient-boosted decision trees~\cite{TreeLUT}.
Another related approach is the family of Weightless Neural Networks (WNNs), which stores learned patterns directly in LUTs~\cite{WISARD, LogicWiSARD, WNN_Edge, WNN_prune, DWN}, though often at the cost of representational power.

Conceptually, KANs are close to PolyLUT~\cite{PolyLUT}, PolyLUT-Add~\cite{PolyLut_Add}, and DWNs \cite{DWN} but possess distinct structural properties. 
While PolyLUT tabulates multivariate polynomials, which theoretically allows the native representation of arbitrary products $p(\mathbf{x}) = \prod_i x_i$, this approach suffers from exponential LUT growth relative to input dimension.
In contrast, KANs decompose functions into sums of tabulated univariate splines. 
Although this formulation relies on layer composition to approximate the multiplicative terms inherent to PolyLUT, it yields linear scaling with input dimension and an additive structure that is naturally amenable to pruning.
While this formulation doesn’t explicitly represent pure multiplicative terms, in practice, compositions of low-order ($\le$3) splines approximate such interactions effectively. Moreover, DWN’s full binarization of inputs and LUTs hinders generalization beyond classification, while \kanele supports higher-precision arithmetic for tasks such as autoencoding and continuous control.
Finally, DWN’s finite-difference differentiability may further constrain optimization flexibility compared to \kanele gradient descent.

\section{KAN Architecture and Quantization-Aware Training and Pruning}
We design the \kanele framework for KAN FPGA deployment using quantization-aware training and pruning, enabling efficient hardware translation while preserving consistency between training and inference.

\subsection{KAN Architecture with Learnable Activation Functions}
Before introducing quantization and pruning, we first outline the core architecture. 
Unlike MLPs, KANs replace fixed nonlinearities with \emph{learnable activation functions}, each modeled as a linear combination of B-spline basis functions with trainable coefficients. 
B-splines are piecewise polynomials defined on a grid, providing smoothness, locality, and efficient nonlinear parameterization. 
More generally, activations can be expressed in other orthogonal bases, such as Fourier series \cite{KAN_GNN, kafn2025, FourierKAN2024}.  

\begin{figure}[]
    \centering
    \includegraphics[width=0.75\linewidth]{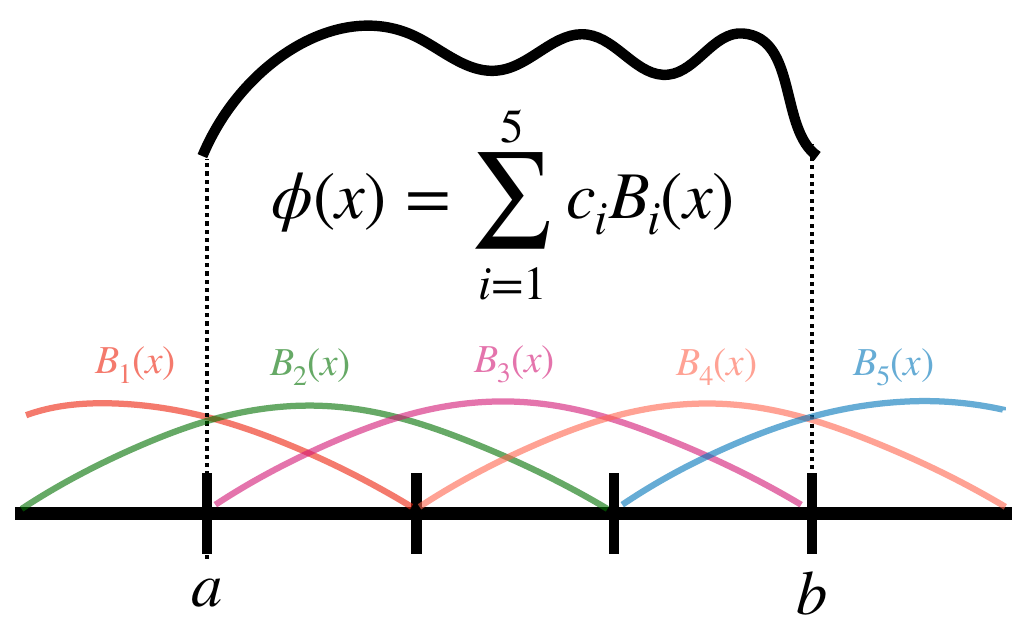}
    \caption{A KAN activation \(\phi(x)\) represented as a linear combination of B\mbox{-}spline basis functions \(B_i(x)\) on a grid over \([a,b]\): 
    \(\phi(x) = \sum_i c_i B_i(x)\). Trainable coefficients \(c_i\) control the overall function shape.}
    \label{fig:learn_act}
\end{figure}

A KAN layer with $d_{\mathrm{in}}$ inputs and $d_{\mathrm{out}}$ outputs is represented as a matrix of 1D learnable functions
\begin{equation}
    \Phi = \{ \phi_{q,p} \}, \quad 
    p = 1,\dots,d_{\mathrm{in}}, \quad 
    q = 1,\dots,d_{\mathrm{out}},
\end{equation}
where each $\phi_{q,p}$ is trainable (Fig.~\ref{fig:learn_act}).

For improved convergence, each $\phi_{q,p}$ combines a base activation $\phi(\cdot)$ with B-splines $\{B_{k}(\cdot)\}$:
\begin{equation}
    \phi_{q,p}(x_p) 
    = w^{\mathrm{base}}_{q,p}\,\phi(x_p) 
    + \sum_{k=1}^{G+S} w^{\mathrm{spline}}_{q,p,k}\, B_{k}(x_p),
\end{equation}
where $w^{\mathrm{base}}_{q,p}$ and $w^{\mathrm{spline}}_{q,p,k}$ are trainable. Splines are defined on a grid of size $G$ and order $S$ within a fixed domain $[a,b]$.  

Given $x_l \in \mathbb{R}^{d_{\mathrm{in}}}$, the output is
\begin{equation}
    (x_{l+1})_j = \sum_{i=1}^{d_l} \phi_{j,i}(x_{l,i}), 
    \quad j = 1,\dots,d_{l+1},
\end{equation}
or in compact form
\begin{equation}
    x_{l+1} = \Phi_l(x_l),
\end{equation}
with $\Phi_l$ the function matrix of layer $l$.  
A $L$-layer KAN is thus
\begin{equation}
    \mathrm{KAN}(x) = \Phi_{L-1} \circ \Phi_{L-2} \circ \cdots \circ \Phi_0 (x).
\end{equation}
As illustrated in Fig.~\ref{fig:teaser}, KANs extend MLPs by learning activations directly, offering greater representational flexibility while preserving a structured, layer-wise graph.

\subsection{Quantization-Aware Training}
\label{sec:QAT}

For efficient FPGA deployment, we adopt quantization-aware training (QAT) via AMD's 
\texttt{Brevitas} library~\cite{brevitas}. Quantizers are placed at the network input and after each KAN layer, ensuring that training adapts to the required hardware precision.

For a layer $l$ with output $\mathbf{x}_{l+1} \in \mathbb{R}^{d_{l+1}}$, the quantized output is
\begin{equation}
    \mathbf{x}_{l+1,q} = q_l(\mathbf{x}_{l+1}),
\end{equation}
where $q_l(\cdot)$ is the layer quantizer. Similarly, an input quantizer $q_I(\cdot)$ is applied to $\mathbf{x}_0$.

\begin{figure}[bp]
    \centering
    \includegraphics[width=0.65\linewidth]{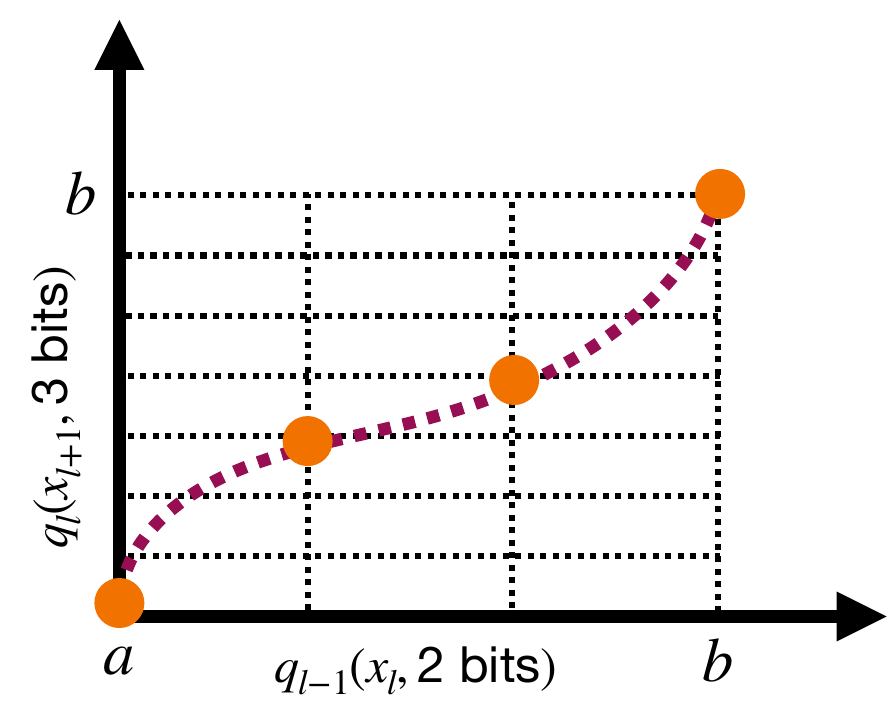}
    \caption{Layer-wise uniform quantization. Here, 2-bit inputs $q_{l-1}(x_l)$ are mapped to 3-bit outputs $q(x_{l+1})$ over the fixed range $[a,b]$. Orange markers indicate quantization levels; the dotted curve is the underlying continuous mapping.}
    \label{fig:lut_quantized}
\end{figure}

The layer output quantizer performs $n_l$-bit uniform quantization:
\begin{equation}
    \mathbf{x}_{l+1,q} 
    = s_l \cdot \text{Quantize}[n_l]\!\left(
        \frac{\operatorname{clip}(\mathbf{x}_{l+1}, a, b)}{s_l}
    \right),
\end{equation}
where $s_l$ is a learnable scale (fixed at inference), and $[a,b]$ is the shared quantization domain (Fig. \ref{fig:lut_quantized}). 

The input quantizer incorporates both scale $s_I$ and bias $b_I$ to handle asymmetric distributions:
\begin{equation}
    \mathbf{x}_{0,q} 
    = s_I \cdot \text{Quantize}[n_I]\!\left(
        \frac{\operatorname{clip}(\mathbf{x}_0, a, b)}{s_I} + b_I
    \right).
\end{equation}
During RTL generation, $s_I$ and $b_I$ are fixed for deterministic behavior.  

In practice, input preprocessing is realized by a batch normalization (zero mean, unit variance) followed by a \texttt{ScalarBiasScale} block introducing $b_I$ and $s_I$. At inference, BN statistics are folded into these constants, yielding an affine shift–scale, clipping, and quantization. This design preserves LUT-based compatibility while avoiding the overhead of full batch normalization.

During training, quantizer gradients are approximated using the straight-through estimator (STE):
\begin{equation}
    \frac{\partial q(x)}{\partial x} \approx 1,
\end{equation}
which allows gradient flow through quantized operations without modification.

\subsection{Pruning via Norm-Based Selection}
\label{sec:prune}

To reduce resource usage, we prune spline connections by evaluating their contribution over the input domain.
In contrast to conventional LUT-based neural networks---which rely on sequential LUT indexing, making every LUT entangled with the next and thus nearly impossible to prune without breaking the model---\textsc{\kanele} exploits the inherently additive structure of KANs, where each LUT contributes independently to a summation.
This independence makes pruning both mathematically natural and directly compatible with FPGA hardware.
The original KAN paper emphasizes efficient pruning as a key advantage of KAN's edge-centric architecture, and we extend this insight to demonstrate a distinct advantage over node-based LUT networks for efficient hardware translation \cite{KAN}.

For each pair $(i,j)$ of input and output neurons, we compute the
activation of the spline component:
\begin{equation}
    f_{p \to q}(x) = \sum_{k=1}^{G+S} w^{\mathrm{spline}}_{q,p,k} \, B_{k}(x).
\end{equation}
Its importance is measured via the $\ell_2$ norm across a sampled input grid $\mathcal{X}$ consistent with its quantization level:
\begin{equation}
    \| f_{p \to q} \|_2 = \left( \sum_{x \in \mathcal{X}} \left| f_{p \to q}(x) \right|^2 \right)^{1/2}.
\end{equation}
A structured pruning mask is then applied:
\begin{equation}
    m_{q,p} =
    \begin{cases}
        1, & \| f_{p \to q} \|_2 > \tau(t), \\
        0, & \text{otherwise},
    \end{cases}
\end{equation}
where $\tau(t)$ is a pruning threshold that changes as a function of epochs ($t$) with $$\tau(t) = T \text{exp}\left(-\ln 20 \cdot \dfrac{\max(t, t_0)}{t_f - t_0}\right).$$ This pruning threshold corresponds to an exponential warmup, where pruning starts on epoch $t_0$ and increases exponentially, hitting $95\%$ of the full pruning threshold $T$ on target epoch $t_f$. This allows us to control pruning dynamics to avoid interference with proper training. Backward pruning is additionally applied if the corresponding output neuron has no active connections in the subsequent layer, ensuring consistent sparsity propagation.

\subsection{KAN Hyperparameters}
\label{sec:kan-hyper}
To summarize, the training and deployment of KANs involves hyperparameters which can be broken up into three main classes: spline representation hyperparameters, hardware architecture hyperparameters, and pruning hyperparameters. 
The descriptions and impact of each hyperparameter are detailed in Table \ref{tab:kan_params}. The joint optimization of these parameters provides a flexible design space that balances learning capacity with hardware efficiency in FPGA deployments.

\begin{table*}[t]
    \centering
    \caption{Summary of KAN training parameters. The first group influences the accuracy through spline representation, the second group encodes the hardware architecture, and the third group determines pruning policy which affects hardware architecture.}
    \label{tab:kan_params}
    \begin{tabularx}{\textwidth}{lXl}
        \toprule
        \textbf{Symbol} & \textbf{Description} & \textbf{Impact} \\
        \midrule
        $G$ & Grid size (number of intervals) & Controls spline resolution; accuracy only \\
        $[a,b]$ & Grid range (spline domain) & Defines support of basis functions; accuracy only \\
        $S$ & Spline order & Smoothness and flexibility; accuracy only \\
        \midrule
        $d_{l}$ & Layer dimensions & Affects model capacity and resource usage \\
        $n_{l}$ & Layer bitwidth (QAT precision) & Direct trade-off between accuracy and resource cost \\
        \midrule
        $T$ & Pruning threshold & Governs sparsity and LUT reduction and accuracy tradeoff \\
        $t_0$ & Warmup start epoch & Determines when pruning warmup starts \\
        $t_f$ & Warmup target epoch & Affects pruning warmup rate
        \\
        \bottomrule
    \end{tabularx}
\end{table*}

\section{LUT-Based KAN Architecture}
\label{sec:lut_kan_fpga}

This section outlines our end-to-end mapping of trained KANs to synthesizable VHDL RTL and associated pipelining strategies.
Our end-to-end toolflow currently supports the basic KAN architecture using B-splines, as in the original paper \cite{KAN}.
Extensions to other bases or architectures such as convolutions or transformers are feasible. The framework is designed for usability—anyone familiar with training MLPs can readily train and deploy \kanele.

\subsection{Toolflow}
\begin{figure}
    \centering
    \includegraphics[width=\linewidth]{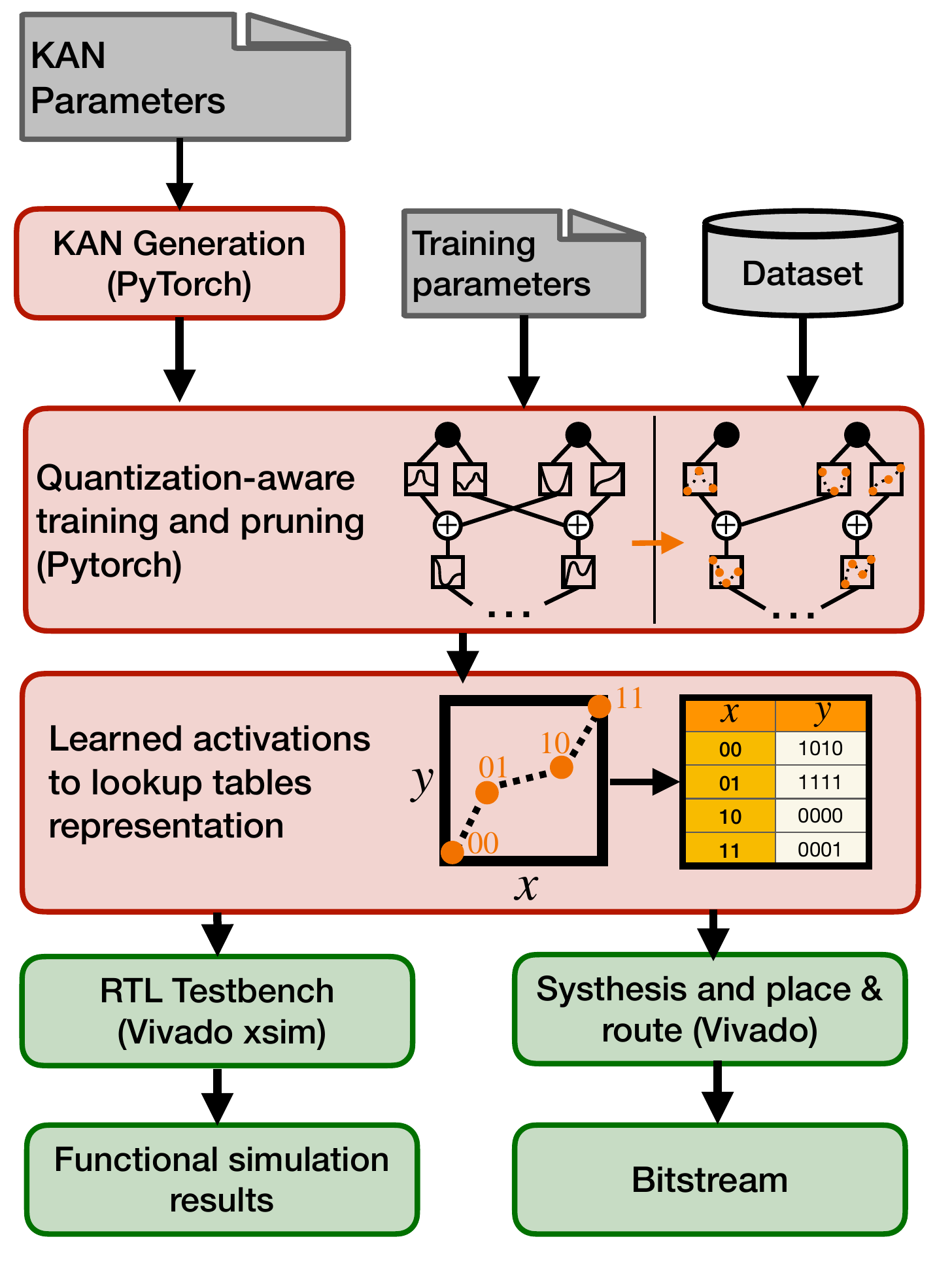}
    \caption{Visualization of the KAN to FPGA implementation toolflow.}
    \label{fig:toolflow}
\end{figure}

A high-level overview of the toolflow stages is shown in Figure \ref{fig:toolflow}.
This push-button workflow removes manual RTL work: starting from a trained PyTorch checkpoint, it deterministically emits RTL, memory images, and build/simulation scripts, enabling rapid deployment of arbitrary KAN topologies to FPGA and bitstream generation via Vivado.

\subsubsection{Training} 
The training process begins by specifying the KAN hyperparameters (see Section~\ref{sec:kan-hyper}) together with the dataset to be used. 
The user may freely select the optimizer and learning-rate schedule best suited to their task. 
In our implementation, we adopt the PyTorch \texttt{AdamW} optimizer~\cite{adamW} as the default choice due to its robustness in handling weight decay. 
The model is then trained with the QAT and pruning mechanism described in Section \ref{sec:QAT} and Section \ref{sec:prune}, respectively, to reduce resource usage and latency, while preserving accuracy. 
This step produces compact learned activations for KAN that can be efficiently mapped to LUT representations for FPGA deployment.

\subsubsection{KAN to Logical-LUTs Conversion}  

Following \cite{NeuralLUT_Assemble}, we denote lookup tables extracted from the network as \emph{Logical-LUTs (L-LUTs)}, and FPGA fabric resources as \emph{Physical-LUTs (P-LUTs)}.
From a trained, pruned, and quantized PyTorch KAN, each surviving connection is translated into an L-LUT.
For every active edge, the input state space is enumerated and the KAN layer’s pre-activation response is evaluated and quantized.
This produces per-connection truth tables stored as compact JSON files, yielding a deterministic, bit-accurate mapping of the model into integer-valued L-LUTs.
The representation preserves quantization and sparsity, enabling efficient FPGA deployment.

\subsubsection{RTL File Generation}  
From the L-LUT graph, we generate a complete RTL design for FPGA deployment.  
The tool emits VHDL sources for the KAN core, per-layer packages, LUT entities, and memory initialization files encoding the truth tables. 
A configuration package specifies bit widths, signal types, and accumulator sizes. 
Each L-LUT is instantiated as a memory-mapped component, organized into layers, with balanced adder trees for output accumulation.
Pipeline registers are inserted between layers to improve clocking and shorten critical paths. 
The result is a self-contained firmware bundle that includes simulation testbenches, initialization vectors, and Vivado build scripts—supporting functional simulation, latency evaluation, and FPGA synthesis.

\subsubsection{Synthesis and Place \& Route}  
We synthesize the generated RTL with \texttt{Vivado~2024.1}, targeting the \texttt{xcvu9p-flgb2104-2-i} FPGA for benchmarking against LUT-based networks, and the \texttt{xczu7ev-ffvc1156-2-e} FPGA for comparison with prior KAN works.  
To ensure fairness and consistency with prior works, we use settings that isolate core delay and area, specifically Vivado’s \texttt{Flow\_PerfOptimized\_high} mode with \texttt{Out-of-Context} synthesis, allowing each module to be compiled independently. While the maximum clock is ultimately limited by the FPGA’s global clock, LUT-centric designs such as \kanele~ typically sustain high frequencies, making the computational core unlikely to be the critical path in larger systems.
The target clock period is therefore chosen relative to network size, following prior work.

\subsection{Pipelining Strategies}
Efficient pipelining is crucial for achieving high FPGA clock frequencies while maintaining low latency across KAN layers. 
We introduce pipelining at two levels: (i) within adder trees that accumulate outputs of multiple Logical-LUTs (L-LUTs) per channel, and (ii) between consecutive network layers.

\begin{figure}
    \centering
    \includegraphics[width=\linewidth]{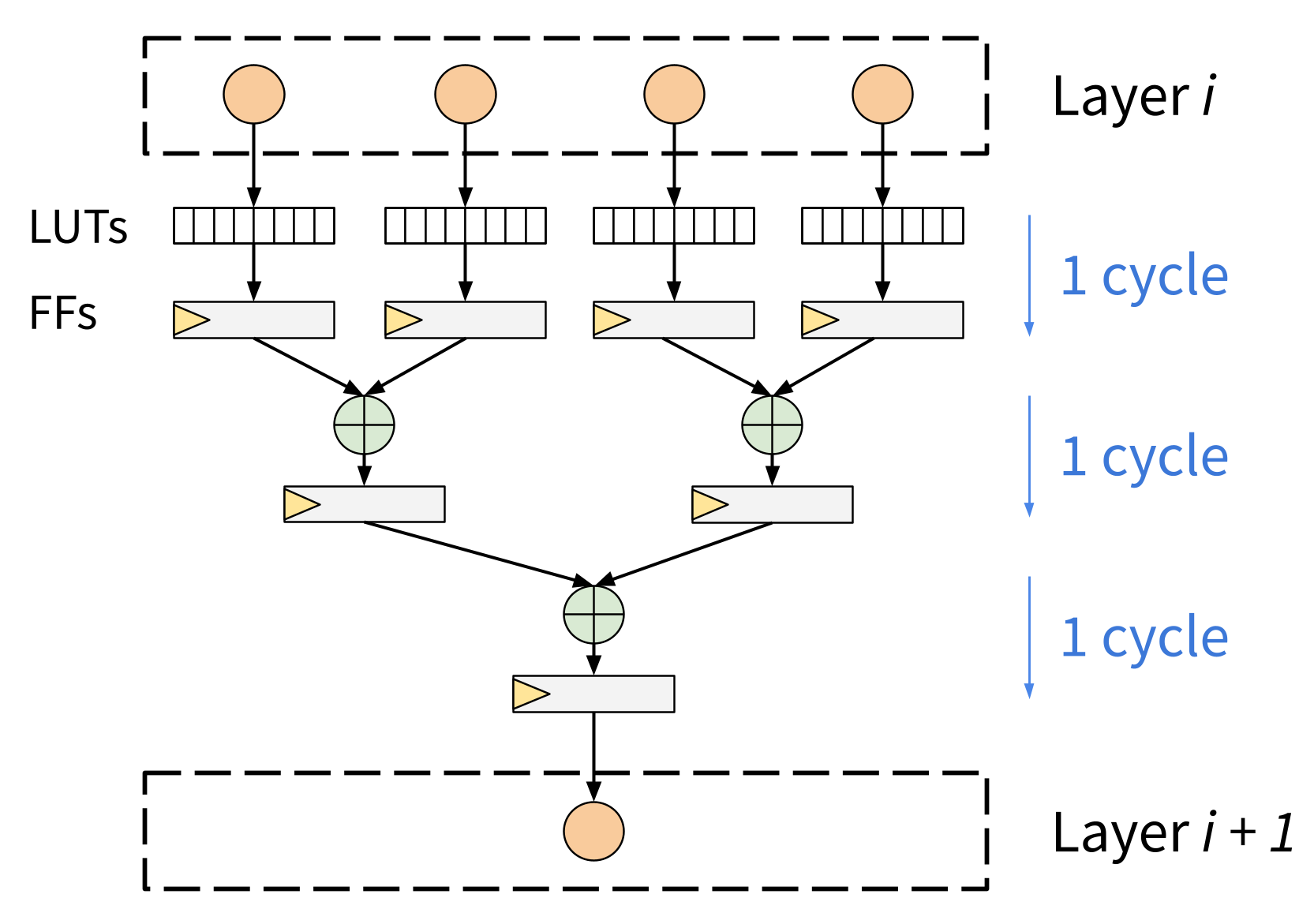}
    \caption{Balanced, pipelined adder tree for computing one neuron activation with $n_{add}=2$.}
    \label{fig:adder_tree}
\end{figure}

\paragraph{Adder Tree Pipelining.} 
Each neuron computes a weighted sum of active inputs via a reduction tree over L-LUT outputs. 
A naïve single-stage sum creates a long combinational path, thereby limiting frequency. 
Instead, we implement a \emph{balanced, pipelined adder tree} with registers after each stage. 
At each stage up to $n_\mathrm{add}$ inputs are combined, reducing fan-in and distributing additions over multiple cycles. 
The depth is
\[
\text{depth}_\ell = \left\lceil \log_{n_\mathrm{add}}(N_\ell) \right\rceil,
\]
as shown in Figure \ref{fig:adder_tree} for $n_\mathrm{add}=2$. 
At the end of the adder tree, quantization and saturation of the sum are performed to make the output consistent with the subsequent layer's input.
This is taken into account during training, preventing any degradation in accuracy.

\paragraph{Inter-Layer Pipelining.}
Pipeline registers are also inserted between layers, capturing saturated outputs before feed-forward. 
This isolates LUT evaluation, summation, and activation in time, minimizing critical paths and balancing latency. 
Register insertion is automated in RTL generation, ensuring deep pipelining for arbitrary KAN topologies.

\paragraph{Limitations.}
As a LUT-based model, KAN inherits known limitations. 
LUT size scales exponentially with input bitwidth, though only linearly with fan-in \cite{PolyLUT, NeuraLUT}. 
For high-dimensional inputs, preserving KAN’s structure requires long adder chains, reducing throughput and increasing resource usage. 
Adder trees sustain high clock frequencies by adding pipeline stages, at the cost of extra clock cycles.
Thus, image tasks like MNIST require aggressive pruning to remain resource-feasible, with some accuracy loss.

\section{Experimental Results}
\subsection{Benchmarks}
We evaluate KAN across three domains of datasets: (i) widely adopted benchmark datasets from the LUT-based neural network literature, (ii) synthetic and tabular datasets previously used in KAN FPGA benchmarking~\cite{Tran_KAN_bad}, and (iii) MLPerf Tiny datasets, which provide real-world tasks with more complex modalities and objectives. 
Together, these domains form a diverse testbed to study KAN across different tasks and levels of complexity.
Below we briefly describe each dataset.

\subsubsection{LUT-based Neural Network Benchmarks}
\begin{itemize}
    \item \textbf{MNIST:} A large-scale handwritten digit recognition dataset containing 60,000 training and 10,000 test images of size $28 \times 28$, labeled across 10 classes (digits 0--9) \cite{MNIST}.
    \item \textbf{JSC OpenML:} A tabular dataset \cite{CERNBox} from the JSC suite, consisting of 16 jet substructure input features and a 5-class jet classification task. It has been widely used in comparisons of LUT-based networks. This version contains 830,000 instances and is known to exhibit easier convergence, possibly due to improved data curation~\cite{NeuralLUT_Assemble}.  

    \item \textbf{JSC CERNBox:} Another dataset \cite{OpenML} from the JSC benchmark, involving the same jet tagging task. It comprises 986,806 instances and is generally considered more challenging, as models trained on it tend to achieve lower accuracies due to the increased dataset complexity.  

\end{itemize}

\subsubsection{KAN FPGA Benchmarks}
\begin{itemize}
    \item \textbf{Moons:} A synthetic two-class dataset \cite{scikit-learn} commonly used for testing nonlinear decision boundaries. Each point lies in one of two interleaving half-moon shapes with added Gaussian noise.
    \item \textbf{Wine:} A dataset from the University of California, Irvine (UCI) Machine Learning Repository \cite{wine_dataset}, containing 13 physicochemical attributes of wine samples classified into 3 quality categories.
    \item \textbf{Dry Bean:} Another UCI dataset \cite{dry_bean} with 16 numerical features representing bean shape and texture, used for classifying 7 different bean varieties .
\end{itemize}

\subsubsection{MLPerf Tiny Benchmarks}
\begin{itemize}
    \item \textbf{ToyADMOS:} An audio anomaly detection dataset featuring sound files from both normally functioning and defective toy cars. An autoencoder is trained on sliding windows of the downsampled mel spectrogram (input size 64). For classification, the mean reconstruction loss across all sliding window spectrograms of an audio file is computed, and a fixed threshold is applied to label the sample as an anomaly. Notably, this benchmark presents significantly more complex inputs compared to the LUT-NN and KAN-FPGA benchmarks while also utilizing a non-classification objective (reconstruction loss). \cite{mlperf}.
\end{itemize}

These datasets span a spectrum of complexity, from low-dimensional toy datasets (Moons) to high-dimensional real-world data (JSC, MNIST, ToyADMOS), ensuring that both the representational capacity and hardware efficiency of \kanele are thoroughly evaluated.

\begin{table*}[htp]
\centering
\caption{Accuracy comparison of MLP Floating Point and KAN Floating Point and Quantized models on benchmark datasets. All models have the same dimensions listed. The floating point versions only use the layer size ($d_l$) parameter.}
\label{tab:mlp-kan-comparison}
\setlength{\tabcolsep}{4pt}
\renewcommand{\arraystretch}{1.1}

\resizebox{\textwidth}{!}{%
\begin{tabular}{l *{6}{c} *{3}{c}}
\toprule
\multirow{2}{*}{\textbf{Datasets}} &
\multirow{2}{*}{$G$} & 
\multirow{2}{*}{$[a,b]$} & 
\multirow{2}{*}{$S$} & 
\multirow{2}{*}{$d_l$} & 
\multirow{2}{*}{$n_l$} &
\multirow{2}{*}{$T$} &
\multicolumn{3}{c}{\textbf{Accuracy (\%)/AUC}} \\
\cmidrule(lr){8-10}
& & & & & & & \textbf{MLP FP} & \textbf{KAN FP} & \textbf{KAN Quantized \& Pruned} \\
\midrule
\textbf{KAN FPGA Benchmarks~\cite{Tran_KAN_bad}} & & & & & & & & & \\
\midrule
Moons      & 6 & [-8, 8]  & 3  & [2, 2, 1]  & [6, 5, 8]  & 0. & 87.2 & \textbf{97.7} & 97.4 \\
Wine       & 6 & [-8, 8] & 3 & [13, 4, 3] & [6, 7, 8] & 0. & 96.3 & 98.1 & \textbf{98.2} \\
Dry Bean   & 6 & [-8, 8] & 3 & [16, 2, 7] & [6, 6, 8] & 0. & 90.9 & \textbf{92.2} & 92.1 \\
\midrule
\textbf{LUT-based neural network benchmarks} & & & & & & & & & \\
\midrule
MNIST       & 30 & [-8,8] & 3 & [784, 62, 10] & [1, 6, 6] &  1. & 96.7 & \textbf{97.9} & 96.3 \\
JSC CERNBox & 30 & [-2, 2]   & 10   & [16, 12, 5]   & [8, 8, 6]   &  0.14  & 73.0 & \textbf{75.1}   & \textbf{75.1} \\
JSC OpenML  & 40 & [-2, 2]    & 10    & [16, 8, 5]    & [6, 7, 6]    & 0.9 & 76.5 & \textbf{76.5} & 76.0 \\
\midrule
\textbf{MLPerf Tiny Benchmark} & & & & & & & & & \\
\midrule
ToyADMOS  & 30 & [-2, 2]    & 10    & [64, 16, 8, 16, 64]    & [7, 8, 8, 7, 8]    & 0.9 & 0.80 & \textbf{0.83} & \textbf{0.83} \\
\bottomrule
\end{tabular}%
}
\end{table*}

\subsection{Training Parameters}
\label{sec:training_params}
KAN hyperparameters are straightforward to manage. The spline-related parameters ($G, [a,b], S$) only affect accuracy and can often be set to robust defaults since they do not impact hardware resources.

In practice, balancing model performance with hardware efficiency centers on tuning three key parameters: the layer dimensions ($d_l$), bitwidth ($n_l$), and pruning threshold ($T$).
These directly control the model's capacity, numerical precision, and sparsity.
As demonstrated in Table \ref{tab:mlp-kan-comparison}, this allows the quantized and pruned KANs to achieve competitive accuracy—even outperforming floating-point versions on datasets like Wine (98.2\%)—while being optimized for an efficient FPGA implementation.

\subsection{Comparison with LUT-NN Architectures}
We benchmarked \kanele against state-of-the-art LUT-based architectures on three datasets: JSC CERNBox, JSC OpenML, and MNIST (Table~\ref{tab:neuralut-comparison}).
It should be noted that, as detailed in Table \ref{tab:mlp-kan-comparison}, we assume the same input bitwidth compared to prior works, which is not the case for DWN \cite{DWN} that uses a thermometer encoding to assign distinct floating-point thresholds to each feature, leading to potentially large overhead. 

\paragraph{JSC CERNBox}
On the more difficult JSC CERNBox dataset, \kanele achieves the highest accuracy (75.1\%), tying with NeuralLUT while also using $18\times$ less LUTs and two orders of magnitude fewer resources.
Compared to the best prior model when considering resources, NeuralLUT-Assemble, we obtain slightly higher accuracy with $1.7\times$ fewer LUTs, over $2.4\times$ higher $F_{\text{max}}$, and the lowest Area$\times$Delay product ($4.1\times 10^{4}$).
In contrast, alternatives such as AmigoLUT, PolyLUT, and LogicNets consume an order of magnitude more LUTs, suffer lower accuracy, or are limited to $F_{\text{max}}$ in the 200--500\,MHz range.
Overall, \kanele sits on the Pareto frontier of accuracy versus efficiency, establishing it as the most efficient solution for this task.

\paragraph{JSC OpenML} On the easier JSC OpenML dataset, most neural networks plateau around 76\% accuracy. 
KAN\mbox{-}LUT reaches this level (76.0\%) while using only 1232 LUTs—the fewest among all models and up to $51\times$ fewer than the hls4ml implementation. 
Compared to NeuralLUT\mbox{-}Assemble, KAN\mbox{-}LUT requires $1.44\times$ fewer LUTs and achieves a slightly higher $F_{\text{max}}$ (987 vs.\ 941\,MHz), though its longer latency (7.1\,ns vs.\ 2.1\,ns) results in a larger Area$\times$Delay. 
Other networks achieve marginally higher accuracy, but only at the cost of substantially greater resource usage and latency. 
Overall, KAN\mbox{-}LUT offers one of the best trade-offs between accuracy and efficiency on this task.
\paragraph{MNIST}
On the MNIST dataset, \kanele achieves a high accuracy of 96.3\%, though some specialized models like NeuraLUT-Assemble and DWN reach close to 98\%.
In terms of hardware resources, DWN is the most compact with 2092 LUTs.
\kanele, with 3809 LUTs, is still significantly more efficient than the majority of other high-accuracy models. For instance, it uses over 20 times fewer LUTs than PolyLUT (75131 LUTs) while achieving only 2\% lower accuracy.
Architectures like NeuraLUT-Assemble and TreeLUT excel in latency (2.1 ns and 2.5 ns) and achieve the best Area$\times$Delay products.
This suggests that the architectural priors of these specialized models may be better aligned with the spatial structure inherent in image data than the function-approximation paradigm of KANs.
Consequently, extending KANELÉ toward convolutional architectures appears to be a promising direction for future work on image-based tasks.

In summary, \kanele consistently demonstrates an exceptional trade-off between predictive accuracy and hardware resource utilization across different benchmarks.
KAN achieves an efficient balance of FPGA resources by leveraging both LUTs and FFs in a complementary manner.
It particularly excels in complex, resource-intensive tasks like the JSC CERNBox benchmark, where it sets a new state-of-the-art in terms of the Area$\times$Delay product.
Based on the nature of these datasets, it can also be inferred that \kanele is better suited for tasks involving symbolic or physical formulas between the input and outputs (e.g., JSC variants), which naturally aligns with the Kolmogorov-Arnold representation theorem.

\begin{table*}[t]
\centering
\caption{Evaluation of \kanele\ against state-of-the-art ultra-low-latency, resource-efficient LUT-based network architectures.
Results are reported after performing \textit{out-of-context} synthesis and place-and-route.
Input bit-widths are consistent with those used in prior works for fair comparison.}
\label{tab:neuralut-comparison}
\resizebox{\textwidth}{!}{
\begin{tabular}{l l r r r r r r r r}
\toprule
\textbf{Dataset} & \textbf{Model} & \textbf{Accuracy (\%)} & \textbf{LUT} & \textbf{FF} & \textbf{DSP} & \textbf{BRAM} & \boldmath{$F_{\max}$} \textbf{(MHz)} & \textbf{Latency (ns)} & \textbf{Area$\times$Delay (LUT$\times$ns)} \\
\midrule
\multirow{6}{*}{\textbf{JSC CERNBox}}
& \textbf{\kanele} & \textbf{75.1} & \textbf{5034} & 1917 & 0 & 0 & \textbf{870} & 8.1 & \textbf{ \boldmath \(4.1\times 10^{4}\)} \\
& NeuraLUT-Assemble~\cite{NeuralLUT_Assemble} & 75.0 & 8539 & 1332 & 0 & 0 & 352 & \textbf{5.7} & \(4.87\times 10^{4}\) \\
& AmigoLUT-NeuraLUT~\cite{AmigoLUT} & 74.4 & 42742 & 4717 & 0 & 0 & 520 & 9.6 & \(4.10\times 10^{5}\) \\
& PolyLUT-Add~\cite{PolyLut_Add} & 75.0 & 36484 & 1209 & 0 & 0 & 315 & 16 & \(5.84\times 10^{5}\) \\
& NeuraLUT~\cite{NeuraLUT} & \textbf{75.1} & 92357 & 4885 & 0 & 0 & 368 & 14 & \(1.29\times 10^{6}\) \\
& PolyLUT~\cite{PolyLUT} & 75.0 & 246071 & 12384 & 0 & 0 & 203 & 25 & \(6.15\times 10^{6}\) \\
& LogicNets~\cite{LogicNets} & 72.0 & 37931 & \textbf{810} & 0 & 0 & 427 & 13 & \(4.93\times 10^{5}\) \\
\midrule
\multirow{4}{*}{\textbf{JSC OpenML}}
& \textbf{\kanele} & 76.0 & \textbf{1232} & 900 & 0 & 0 & \textbf{987} & 7.1 &   $8.7\times 10^{3}$  \\
& NeuraLUT-Assemble~\cite{NeuralLUT_Assemble} & 76.0 & 1780 & 540 & 0 & 0 & 941 & \textbf{2.1} & \boldmath{\(3.92\times 10^{3}\)} \\
& TreeLUT~\cite{TreeLUT} & 75.6 & 2234 & \textbf{347} & 0 & 0 & 735 & 2.7 & \(6.03\times 10^{3}\) \\
& DWN~\cite{DWN} & 76.3 & 4972 & 3305 & 0 & 0 & 827 & 7.3 & \(3.6\times 10^{4}\) \\
& da4ml~\cite{da4ml} & \textbf{76.9} & 12250 &  1502 & 0 & 0 & 212 & 18.9 & \(2.3\times 10^{5}\) \\
& hls4ml (Fahim et al.)~\cite{hls4ml_Fahim} & 76.2 & 63251 & 4394 & 38 & 0 & 200 & 45 & \(2.85\times 10^{6}\) \\
\midrule
\multirow{10}{*}{\textbf{MNIST}} 
& \textbf{\kanele} & 96.3 & 3809 & 4133 & 0 & 0 & 864 & 9.3 & $3.5 \times 10^4$ \\
& NeuraLUT-Assemble~\cite{NeuralLUT_Assemble} & \textbf{97.9} & 5070 & 725 & 0 & 0 & 863 & \textbf{2.1} & \boldmath{$1.06\times 10^{4}$} \\
& TreeLUT~\cite{TreeLUT} & 96.6 & 4478 & \textbf{597} & 0 & 0 & 791 & 2.5 & \(1.12\times 10^{4}\) \\
& DWN~\cite{DWN} & 97.8 & \textbf{2092} & 1757 & 0 & 0 & 873 & 9.2 & \(1.92\times 10^{4}\) \\
& PolyLUT-Add~\cite{PolyLut_Add} & 96.0 & 14810 & 2609 & 0 & 0 & 625 & 10 & \(1.48\times 10^{5}\) \\
& AmigoLUT-NeuraLUT~\cite{AmigoLUT} & 95.5 & 16081 & 13292 & 0 & 0 & \textbf{925} & 7.6 & \(1.22\times 10^{5}\) \\
& NeuraLUT~\cite{NeuraLUT} & 96.0 & 54798 & 3757 & 0 & 0 & 431 & 12 & \(6.58\times 10^{5}\) \\
& PolyLUT~\cite{PolyLUT} & 97.5 & 75131 & 4668 & 0 & 0 & 353 & 17 & \(1.38\times 10^{6}\) \\
& FINN~\cite{FINN} & 96.0 & 91131 & --- & 0 & 5 & 200 & 310 & \(2.82\times 10^{7}\) \\
& hls4ml (Ngadiuba et al.)~\cite{hls4ml_Ngadiuba} & 95.0 & 260092 & 165513 & 0 & 345 & 200 & 190 & \(4.94\times 10^{7}\) \\

\bottomrule
\end{tabular}
}
\end{table*}

\begin{table*}[t]
\centering
\caption{FPGA resource utilization and latency of KAN models on Moons, Wine, and Dry Bean benchmarks as used in \cite{Tran_KAN_bad}}
\label{tab:kan-fpga-benchmarks}
\small
\resizebox{\textwidth}{!}{%
\begin{tabular}{l l r r r r r r r r r}
\toprule
\textbf{Dataset} & \textbf{Model} & \textbf{Accuracy (\%)} & \boldmath{$F_{\max}$} \textbf{(MHz)} & \textbf{BRAM} & \textbf{DSP} & \textbf{FF} & \textbf{LUT} & \textbf{Latency (cycles)} & \textbf{Latency (ns)} & \textbf{Area$\times$Delay (LUT $\times$ ns)} \\
\midrule
\multirow{1}{*}{\textbf{Moons}} 
& \textbf{\kanele} & 97 & \textbf{1736} & \textbf{0} & \textbf{0} & \textbf{57} & \textbf{67} & \textbf{5} & \textbf{2.9} & \textbf{\boldmath $1.9 \times 10^2$} \\
& KAN (Tran et al)~\cite{Tran_KAN_bad} & 97 & - & 10 & 120 & 8622 & 17877 & 128 & 1280 &  $2.3 \times 10^7$ \\
& ChebyUnit~\cite{cheby_kan} & \textbf{100}  & - & 10 & 40 & 12150 & 9888 & 13 & 130 & $1.3 \times 10^6$  \\
\midrule
\multirow{1}{*}{\textbf{Wine}} 
& \textbf{\kanele} & \textbf{98} & \textbf{983} & \textbf{0} & \textbf{0} & \textbf{686} & \textbf{534} & \textbf{6} & \textbf{6.1} & \textbf{\boldmath$8.8 \times 10^3$} \\
& KAN (Tran et al)~\cite{Tran_KAN_bad}  & 97 & - & 132 & 950 & 74741 & 146843 & 688 & 6880 & $1.0 \times 10^9$ \\
& ChebyUnit~\cite{cheby_kan} & 95  & - & 132 & 324 & 22104 & 30154 & 13 & 130 & $3.9 \times 10^6$  \\
\midrule
\multirow{1}{*}{\textbf{Dry Bean}} 
& \textbf{\kanele} & \textbf{92} & \textbf{842} & \textbf{0} & \textbf{0} & \textbf{471} & \textbf{402} & \textbf{6} & \textbf{7.1} & \textbf{\boldmath{$3.3 \times 10^3$}} \\
& KAN (Tran et al)~\cite{Tran_KAN_bad}  & \textbf{92} & - & 781 & 9111 & 734544 & 1677558 & 1896 & 18960 &  $3.2 \times 10^{10}$ \\
& ChebyUnit~\cite{cheby_kan} & \textbf{92}  & - & 781 & 256 & 25198 & 27359 & 13 & 130 & $3.6 \times 10^6$  \\
\bottomrule
\end{tabular}%
}
\end{table*}

\begin{table*}[t]
\centering
\caption{Comparison of FPGA resource utilization, latency, and power consumption for the anomaly detection task on the ToyADMOS time series dataset in the MLPerf Tiny Benchmark, evaluated on the \texttt{xc7a100t-1csg324} FPGA \cite{mlperf}.}
\label{tab:tiny-ml-benchmark}
\small
\resizebox{\textwidth}{!}{%
\begin{tabular}{l l r r r r r r r r r r}
\toprule
\textbf{Dataset} & \textbf{Model} & \textbf{AUC} & \textbf{BRAM (36kb)} & \textbf{DSP} & \textbf{FF} & \textbf{LUT} & \textbf{LUTRAM}& \textbf{II (clocks)}& \textbf{Throughput (inf/s)} & \textbf{Latency ($\mu s$)} & \textbf{Energy/inf. ($\mu J$)} \\
\midrule
\multirow{1}{*}{\textbf{ToyADMOS}} 
& \textbf{\kanele} & \textbf{0.83} & \textbf{0} & \textbf{0} & \textbf{17,643} & \textbf{29,981} & \textbf{0} & \textbf{1} &  \textbf{228 M} & \textbf{0.07} & \textbf{0.01} \\
& \texttt{hls4ml} (MLPerf Tiny v0.7)~\cite{Borras_hls4ml} & \textbf{0.83} & 22.5 & 207 & 61,639 & 51,429 & 5,780 & 144 &  694 k & 45 & 98.4 \\
\bottomrule
\end{tabular}%
}
\end{table*}

\subsection{Comparison with Prior KAN-FPGA Literature}
We benchmarked \kanele against the KAN FPGA implementation by Tran et al.~\cite{Tran_KAN_bad} on the Moons, Wine, and Dry Bean datasets. The results, detailed in Table~\ref{tab:kan-fpga-benchmarks}, demonstrate that \kanele offers a dramatic improvement in hardware efficiency and performance while maintaining or exceeding the accuracy of the previous work.

Through its LUT-based approach, our implementation completely eliminates the need for BRAM and DSP blocks, which are heavily utilized in the implementation by Tran et al. This leads to a massive reduction in the overall hardware footprint. For instance, on the Dry Bean dataset, \kanele uses only 402 LUTs and 471 FFs, whereas the previous work consumes over 1.6 million LUTs and 734,000 FFs---a reduction of more than 4000x in LUTs.

\kanele also achieves high maximum frequencies (up to 1736 MHz) and very low latencies. On the Dry Bean benchmark, for example, our model's latency is 7.1 ns, a speedup of over 2600x compared to the 18,960 ns reported by Tran et al. Consequently, \kanele achieves an excellent Area$\times$Delay product across all benchmarks. 

\begin{figure*}[htbp] 
    \centering
    \includegraphics[width=\textwidth]{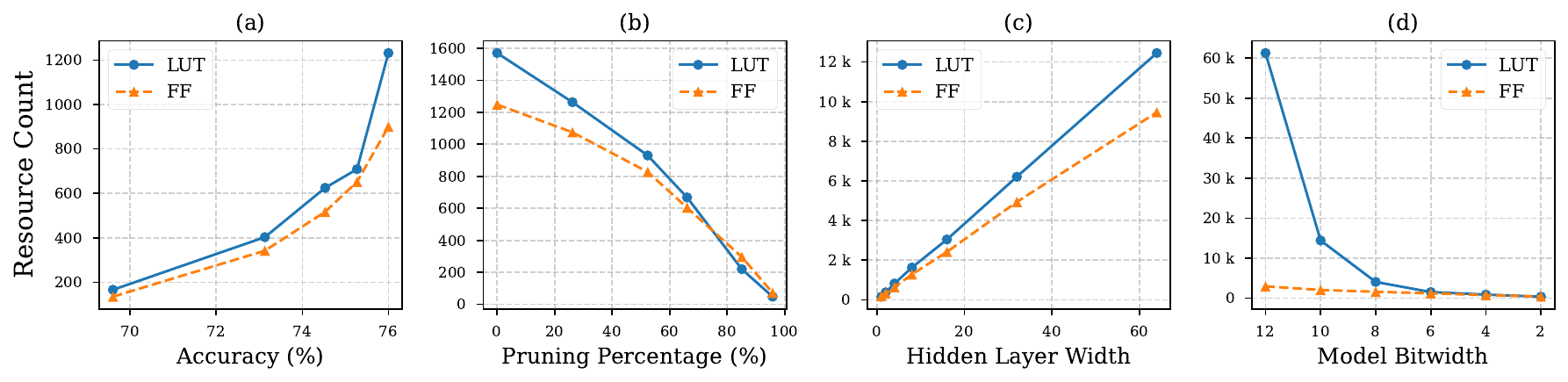}
    \caption{Ablation study of \kanele on the JSC OpenML benchmark demonstrating trade-offs between accuracy, pruning, and resource usage.
    (a) Accuracy improves steadily as hardware resources scale up with LUT and FF usage growing at roughly the same rate. 
    (b) LUT/FF usage scales roughly linearly with the number of unpruned edges.
    (c) LUT/FF usage scales linearly with hidden layer width, confirming a direct mapping from learned activation functions (edges) to resources.
    (d) Decreasing activation bitwidths reduces LUT resource usage exponentially, with diminishing returns observed below 6 bits.}
    \label{fig:ablation}
\end{figure*}

\subsection{Comparison with \texttt{hls4ml} MLPerf Tiny}
To understand the performance of \kanele on more complex datasets, we compare the framework to \texttt{hls4ml} \cite{Borras_hls4ml} on the ToyADMOS dataset, part of the MLPerf Tiny benchmark suite.
Other LUT-based neural networks have not attempted this benchmark, possibly due to its high complexity and/or non-classification-based training scheme.
The results in Table~\ref{tab:tiny-ml-benchmark} indicate that \kanele achieves substantially better performance than prior approaches in terms of both resource efficiency, latency, and power. These improvements highlight the potential of \kanele for future studies involving more complex datasets as well as tasks beyond classification. Specifically, \kanele eliminates the need for BRAM, LUTRAM, and DSPs, while reducing LUT usage by 41.7\% and FF usage by 71.4\% relative to \texttt{hls4ml}. 
In terms of performance, \kanele delivers $330\times$ higher throughput, $643\times$ lower latency and a $9,840\times$ reduction in energy per inference. 
These results establish \kanele as a highly efficient alternative to existing FPGA neural implementations, with strong potential for scaling to more complex datasets and tasks beyond classification.

\subsection{Ablation Study}

We perform an ablation study for \kanele using the JSC OpenML dataset, shown in Figure~\ref{fig:ablation}.
This analysis isolates the effect of four key design factors—accuracy, pruning, hidden layer width, and model bitwidth—on FPGA resource utilization (LUTs and FFs).
It is seen that pruning and quantization provide the most effective levers for controlling hardware footprint, while hidden layer width and accuracy tuning allow fine-grained trade-offs between performance and efficiency.
Another important conclusion is that the size of a KAN network can holistically be thought of in terms of its number of edges: this is proportional to the number of LUTs and FFs used, as seen in Figures ~\ref{fig:ablation}(b) and ~\ref{fig:ablation}(c). Together, these insights guide principled design choices for deploying \kanele under tight FPGA resource budgets.

\subsection{Extension to Real-time Control Systems}
\label{sec:control-system}
To demonstrate that the \kanele paradigm extends well beyond the traditional supervised learning tasks typically studied in the LUT-based neural network community, we apply our framework to a reinforcement learning benchmark.
Specifically, we evaluate on the \texttt{HalfCheetah} environment, a classic continuous control task in reinforcement learning (RL), most commonly accessed through the MuJoCo physics simulator via OpenAI Gym (now Gymnasium) \cite{Mujoco}.
The objective is to learn a policy that enables a simulated two-legged agent to run as fast and as stably as possible.
This environment is widely used as a standard benchmark to compare RL algorithms and function approximators.
While HalfCheetah is a simulated benchmark, it captures key aspects of many practical robot control tasks, as they all rely on the same underlying design and physics principles.

This extension is motivated by prior work such as \cite{KAN_RL}, where the authors showed that KAN actor networks with significantly fewer trainable parameters achieved higher rewards compared to much larger MLPs when trained with the Proximal Policy Optimization (PPO) algorithm.
PPO is one of the most widely used RL algorithms in practice, particularly in robotics, games, and continuous control tasks.

\subsubsection{Setup}
Since the primary focus of this paper is inference efficiency, we incorporated KANs only as the function approximator for the policy (actor), as this is the component that must be deployed in practice.
The value (critic) function remained an MLP for all experiments.
We evaluated four training scenarios: 
\begin{enumerate}
    \item MLP actor + MLP critic (both FP),
    \item Quantized MLP actor (8-bit) + MLP FP critic,
    \item KAN FP actor + MLP FP critic,
    \item Quantized KAN actor (8-bit) + MLP FP critic.
\end{enumerate}

To ensure robustness against variance in random initialization, each scenario was trained with 5 different random seeds, for 1 million environment steps per seed.
The actor networks were chosen such that the MLP actor had roughly five times more trainable parameters than the KAN actor (Table \ref{tab:RL_parameters}),  highlighting that KAN can be more effective even with significantly fewer parameters.

\begin{table}[ht]
\centering
\caption{Network architectures of the actor and critic.}
\resizebox{\columnwidth}{!}{%
\begin{tabular}{lp{0.28\columnwidth}p{0.45\columnwidth}}
\hline
\textbf{Model} & \textbf{Dimensions} & \textbf{Trainable Parameters} \\
\hline
MLP Actor & [17, 64, 64, 6] & 5383 \\
MLP Critic & [17, 64, 64, 6] & 5383 \\
KAN Actor & [17, 6] & 1020 \\
\hline
\end{tabular}%
}
\label{tab:RL_parameters}
\end{table}

\subsubsection{Training results}
\begin{figure}
    \centering
    \includegraphics[width=\linewidth]{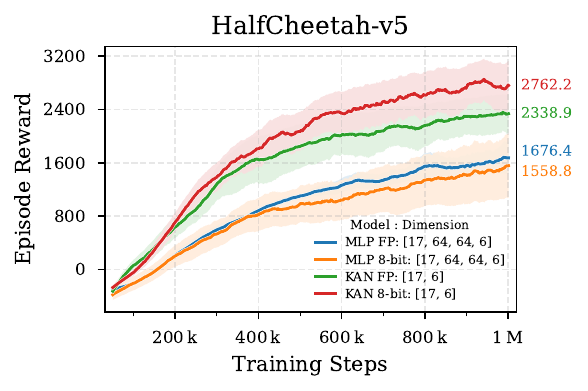}
    \caption{PPO training on \texttt{HalfCheetah-v5} with 5 seeds.
    The quantized KAN actor (8-bit) outperforms both the  KAN (FP) and the larger MLP (FP), despite using $\sim$5$\times$ fewer parameters, showing robustness to quantization and strong parameter efficiency.}
    \label{fig:HalfCheetah}
\end{figure}

Figure~\ref{fig:HalfCheetah} shows the learning curves across the four scenarios.
The quantized KAN actor achieves an average return of \textbf{2762.2}, outperforming both the larger MLP FP and quantized actor baselines (\textbf{1676.4} and \textbf{1558.8}) and the full-precision KAN actor (\textbf{2338.9}).
These results demonstrate that KANs not only remain robust under aggressive 8-bit quantization, but can also benefit from it, potentially due to regularization effects by fixed bit operations.

\subsubsection{Hardware Performance}
To evaluate the efficiency of KANs in practical deployment, we compare the hardware cost of the 8-bit quantized KAN actor against the 8-bit quantized MLP actor.
The KAN actor was deployed using the \kanele framework, while the MLP actor was implemented with \texttt{hls4ml} \cite{hls4ml_Fahim} using the \texttt{Resource} strategy for a fair baseline.
We planned to have both models synthesized on a Xilinx \texttt{xczu7ev-ffvc1156-2-e} FPGA in out-of-context mode.
However, the 8-bit MLP design exceeds the available FPGA resources, so its performance results are based on HLS estimates rather than the actual implementation.
Table~\ref{tab:kan_RL_hardware} summarizes the results.
As can be seen, the 8-bit KAN actor achieves significantly lower resource utilization, latency, and power compared to the 8-bit MLP actor, underscoring the advantages of KANs for real-time reinforcement learning control tasks.

In conclusion, the \texttt{HalfCheetah} task is a simulated environment that captures core principles of real-world control tasks and serves as a strong proxy for domains where real-time, resource-efficient policies are essential.
These results therefore highlight the suitability of KANs for deployment in settings where such constraints matter (e.g., robotics, embedded control, trading, or quantum computing).
Thus, the benchmark serves as a proof-of-concept: KANs can provide competitive or superior RL performance while being dramatically more efficient in terms of size and quantization tolerance.

\begin{table}[t]
\small 
\centering
\caption{FPGA performance of the KAN 8-bit model on the \texttt{HalfCheetah} RL task targeting the \texttt{xczu7ev-ffvc1156-2-e} FPGA. The MLP 8-bit actor does not fit on the FPGA, so its power estimate is unavailable and resource usage is reported from HLS estimates, while KAN 8-bit results are obtained after place-and-route in \texttt{out-of-context} mode.}
\label{tab:kan_RL_hardware}
\begin{tabularx}{\columnwidth}{l *{2}{>{\raggedleft\arraybackslash}X}}
\toprule
\textbf{Metric} & \textbf{KAN 8-bit} & \textbf{MLP 8-bit \texttt{hls4ml}} \\
\midrule
1M Episode Reward  & \textbf{2762.2} & 1558.8 \\
\midrule
Max. Frequency ($F_{\max}$) & \textbf{884 MHz} & 500 MHz \\
Latency & \textbf{4.5 ns} & 893 ns \\
\midrule
BRAM & 0 & 0 \\
DSP & \textbf{0} & 14,346 \\
Flip-Flops (FF) & \textbf{2,828} & 460,800 \\
Look-Up Tables (LUT) & \textbf{1,136} & 230,400 \\
\midrule
Area$\times$Delay & \textbf{\boldmath{$1.3 \times 10^4$} LUT$\cdot$ns} & $2.1 \times 10^8$ LUT$\cdot$ns \\
Dynamic Power & \textbf{0.224 nJ/sample} & $\gg 0.224$ nJ/sample \\
\bottomrule
\end{tabularx}
\end{table}

\section{Conclusion and Future Works}

We present \kanele, a hardware–software co-design framework that maps Kolmogorov–Arnold Networks (KANs) onto a LUT-native computational architecture for FPGAs.
Unlike most existing ML hardware–software co-design approaches, KANs are built entirely from learnable 1D activation functions defined on a fixed domain.
Each learned activation function $\phi(x)$ is thus not merely approximated by a L-LUT: it \emph{is} a lookup table.
Moreover, the additive structure of KANs enables an especially natural form of pruning: each node can be directly removed from the summation without disrupting the remaining computation.  
This is in stark contrast to conventional LUT-based neural networks, where LUTs are typically chained together as indices into one another, making the removal of even a single LUT practically impossible without breaking the model. 
Consequently, mapping a KAN to an FPGA is less a process of compilation and more one of direct instantiation.
This paradigm shift---from emulating arithmetic to directly configuring logic---is what unlocks the extreme efficiency of \textsc{KANELÉ}, sidestepping the need for DSPs and BRAMs entirely and aligning the algorithm directly with the hardware's native capabilities.

Across standard LUT–NN benchmarks and prior KAN–FPGA tasks, \kanele demonstrates strong performance with a favorable Area$\times$Delay trade-off, offering substantially lower latency and reduced logic utilization compared to earlier KAN-on-FPGA designs, while matching or exceeding other LUT-centric architectures when the target function exhibits symbolic or physics-inspired structure.
We further demonstrated the applicability of \kanele to real-time control by deploying an 8-bit KAN policy that surpasses a larger MLP while achieving ultra-low latency and higher resource efficiency on FPGA.

\paragraph{Future Works.}
Building on our results, we identify several avenues to broaden the scope and impact of \kanele:

\begin{itemize}
    \item \textbf{Broader model families.}  
    Extending beyond single KANs to ensembles, temporal and convolutional KANs, graph-based KANs, or transformer-style KANs (``KAN-GPT'') on FPGAs.
    
    \item \textbf{Alternative orthogonal bases.}  
    Moving beyond B-splines by exploring Fourier, wavelet, or rational bases for learnable activations.
    These alternatives may improve approximation power and training dynamics while remaining LUT-compatible.
    
    \item \textbf{Practical deployment in control tasks.}  
    Our reinforcement learning demo shows that 8-bit KAN policies can outperform larger MLPs while sustaining ultra-low FPGA latency, meeting demands of applications such as quantum error correction, plasma stabilization, adaptive optics, and robotics.  
    Future work targets rapid in-field adaptation through hot-swapping edge tables via partial reconfiguration or LUT updates, enabling lightweight online learning with minimal latency.  

\end{itemize} 

Realizing the full potential of this paradigm, however, first requires overcoming the perception that KANs are inherently inefficient in hardware.
Our work, \kanele, directly refutes this view by aligning the activation-centric KAN formulation with the native strengths of FPGA LUT fabrics.
This approach transforms KANs into a practical, high-throughput, and power-efficient inference architecture.
We believe this hardware-centric perspective not only solves a critical implementation challenge but also solidifies the promising path toward interpretable, low-resource neural networks that scale from embedded controllers to large scientific instruments, naturally benefiting from software–hardware co-design.

\begin{acks}
The authors thank the anonymous referees, Vladimir Loncar, Marta Andronic and Ryan Kastner for their valuable comments and helpful suggestions. PH, DH, and AG are supported by the NSF-funded A3D3 Institute (NSF-PHY-2117997).
\end{acks}

\bibliographystyle{ACM-Reference-Format}
\bibliography{bib/references}


\end{document}